\documentclass[prb,showpacs,twocolumn,superscriptaddress,floatfix,longbibliography,nofootinbib]{revtex4-1}
\usepackage{graphicx,amsfonts,amssymb,amsmath,xspace,dsfont,pxfonts}
\usepackage[colorlinks=true,citecolor=green,linkcolor=red,urlcolor=blue]{hyperref}
\usepackage[T1]{fontenc}
\usepackage{lmodern}

\usepackage{color}
\definecolor{g}{rgb}{.1,0.4,.1} % {.0,0.7,.5}
\definecolor{b}{rgb}{0,0.2,1}
\definecolor{rouge}{rgb}{0.82,0.,0.}
\definecolor{vert}{rgb}{0.,0.82,0.}
\definecolor{orange}{rgb}{1,0.5,0.}
\definecolor{bleu}{rgb}{0.,0.,0.82}
\definecolor{m}{rgb}{0.82,0.,0.82}
\definecolor{vert2}{rgb}{0.,0.5,0.}
\definecolor{rougeclair}{rgb}{1.0,0.7,0.7}

\newcommand{\beq}{\begin{equation}}
\newcommand{\be}{\begin{equation}}
\newcommand{\beqn}{\begin{eqnarray}}
\newcommand{\eeq}{\end{equation}}
\newcommand{\ee}{\end{equation}}
\newcommand{\eeqn}{\end{eqnarray}}

\newcommand{\bem}{\begin{pmatrix}}
\newcommand{\eem}{\end{pmatrix}}

\newcommand{\e}{\textrm{e}}
\newcommand{\im}{\textrm{i}}
\newcommand{\dd}{\textrm{d}}

\newlength{\ldag}
\begin{document}

\title{Landau level broadening, hyperuniformity, and discrete scale invariance}

\author{Jean-No\"el Fuchs}
\email{fuchs@lptmc.jussieu.fr}

\author{R\'emy Mosseri}
\email{remy.mosseri@upmc.fr}

\author{Julien Vidal}
\email{vidal@lptmc.jussieu.fr}
\affiliation{Sorbonne Universit\'e, CNRS, Laboratoire de Physique Th\'eorique de la Mati\`ere Condens\'ee, LPTMC, F-75005 Paris, France}

%%%%%%%%%%%%%%%%%%%%%%%%%%%%%%%%%
%%%%%%%%%%%%%%%%%%%%%%%%%%%%%%%%%
\begin{abstract}
We study the energy spectrum of a two-dimensional electron in the presence of both a perpendicular magnetic field and a potential. In the limit where the potential is small compared to the Landau level spacing, we show that the broadening of Landau levels is simply expressed in terms of the structure factor of the potential. For potentials that are either periodic or random, we recover known results. Interestingly, for potentials with a dense Fourier spectrum made of Bragg peaks (as found, e.g., in quasicrystals), we find an algebraic broadening with the magnetic field characterized by the hyperuniformity exponent of the potential. Furthermore, if the potential is self-similar such that its structure factor has a discrete scale invariance, the broadening displays log-periodic oscillations together with an algebraic envelope. 
\end{abstract}

\pacs{}

\maketitle
%%%%%%%%%%%%%%%%%%%%%%%%%%%%%%%%%
%%%%%%%%%%%%%%%%%%%%%%%%%%%%%%%%%

%
%
%%%%%%%%%%%%%%%%%%%%%%%%%%%%%%%%%
%%%%%%%%%%%%%%%%%%%%%%%%%%%%%%%%%
\section{Introduction}
%%%%%%%%%%%%%%%%%%%%%%%%%%%%%%%%%
%%%%%%%%%%%%%%%%%%%%%%%%%%%%%%%%%
%
%

In the presence of a magnetic field, the energy spectrum of noninteracting electrons in two dimensions is known to consist of Landau levels. These discrete energy levels are responsible for many remarkable phenomena, among which is the celebrated integer quantum Hall effect \cite{Klitzing80,Klitzing86}. Each Landau level has a macroscopic degeneracy that is proportional to the strength of the magnetic field. This degeneracy is expected to be lifted by a generic  perturbation leading to a broadening of Landau levels that may have important physical consequences. For instance, plateaus observed in the Hall resistance are directly related to the broadening induced by disorder, as realized early by Ando et al. \cite{Ando74_1,Ando74_2,Ando74_3}. 
Most studies on Landau level broadening focused on disordered systems (see Ref.~\onlinecite{Huckestein95} for a review), but the role played by periodic potentials has also attracted much attention following the original work of Rauh \cite{Rauh74,Rauh75}. Based on a free-electron picture, Rauh's approach also allows one to qualitatively understand the Landau level broadening in the small-field limit of the Hofstadter butterfly for periodic lattices \cite{Hofstadter76,Claro79}, although a quantitative analysis requires a semi-classical treatment \cite{Wilkinson84}.  Recently, the Hofstadter butterfly of some quasiperiodic systems has been investigated, unveiling an unusual broadening of Landau levels \cite{Fuchs18} different from the one expected for periodic or disordered systems, hence suggesting a nontrivial mechanism for potentials with a dense set of Bragg peaks.  

\begin{figure}
\includegraphics[width=0.9\columnwidth]{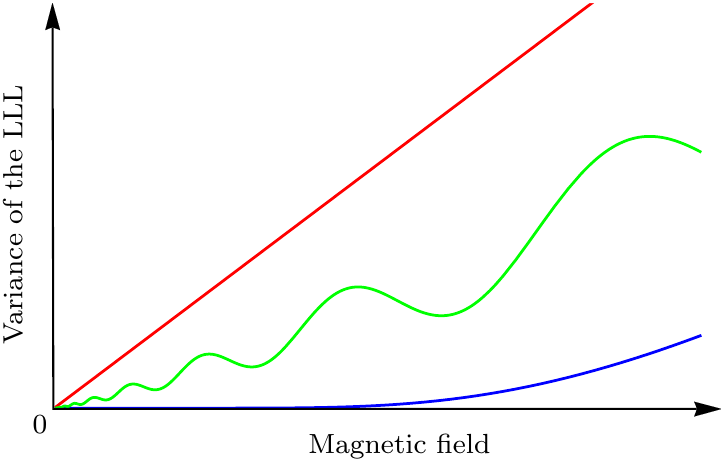}
\caption{Sketch of the LLL variance $w^2$ as a function of the magnetic field $B$ for  different perturbing potentials. Red: disordered \mbox{$w^2\sim B$} (see Ref.~\onlinecite{Ando74_1}). Green: hyperuniform with discrete scale invariance \mbox{$w^2\sim B^\frac{2+\alpha}{2} + \log$-periodic oscillations}, where $\alpha$ is the hyper\-uniformity exponent (this work). Blue: periodic  $w^2\sim {\rm e}^{-\#/B}$ (see Ref.~\onlinecite{Rauh75}).
}
\label{fig:main}
\end{figure}

The goal of the present paper is to provide a general framework to compute the broadening of Landau levels in the presence of an arbitrary potential.  Our main result, given in Eq.~(\ref{LLLvar}), relates the variance of the lowest Landau level (LLL) to the structure factor of the perturbing potential (an extension to higher-energy Landau levels is straightforward). This simple expression reproduces the aforementioned results for disordered and periodic cases, but it also allows us to investigate more subtle potentials (see Fig.~\ref{fig:main} for a summary of the results). In particular, we find that when the Fourier spectrum of the potential is dense and made of Bragg peaks (as in quasicrystals), the variance of the LLL increases algebraically with the magnetic field [see Eq.~(\ref{eq:main_result})] with an exponent characterizing the hyperuniformity of the potential. This notion of hyperuniformity is commonly used to describe sets of points with an unusually large suppression of density fluctuations at long wavelengths \cite{Torquato18}. We also show that if the potential has a discrete scale invariance \cite{Sornette98,Akkermans13}, then the variance displays log-periodic oscillations together with a power-law envelope.  To illustrate these results, we consider three examples of quasiperiodic potentials, for which we compute exactly the hyperuniformity exponent and the period of these oscillations, when it exists. 

\

%
%
%%%%%%%%%%%%%%%%%%%%%%%%%%%%%%%%%
%%%%%%%%%%%%%%%%%%%%%%%%%%%%%%%%%
\section{Landau levels perturbed by a potential}
%%%%%%%%%%%%%%%%%%%%%%%%%%%%%%%%%
%%%%%%%%%%%%%%%%%%%%%%%%%%%%%%%%%
%
%
To begin with, let us recall a few well-known results. The Hamiltonian describing a particle of mass $m$ and charge $e$ in a magnetic field $\mathbf{B}=\boldsymbol{\nabla}\times \mathbf{A}$ is given by
%
%
%%%%%%%%%%%%%%%
\be
H_0=\frac{(\mathbf{p}-e\mathbf{A})^2}{2m}.
\ee
%%%%%%%%%%%%%%%
%
%
Here, we consider a two-dimensional system with a magnetic field perpendicular to the plane. Such a field can be described by the symmetric gauge where the vector potential reads $\mathbf{A}=B(-y/2,x/2,0)$. The spectrum of $H_0$ consists in equidistant energy levels known as Landau levels,
%
%
%%%%%%%%%%%%%%%
\be
E_n=\hbar \omega_{\mathrm c} (n+1/2),  \forall n \in \mathbb{N},
\ee
%%%%%%%%%%%%%%%
%
%
 where $\omega_{\mathrm c}=|eB|/m$ is the cyclotron frequency.  Each Landau level has a degeneracy proportional to the sample area $\mathcal{A}$ and the magnetic field. In the following, we set $\hbar=e=1$.
 
Our aim is to study the behavior of the Landau levels in the presence of a time-independent potential.  Thus, we consider the following general Hamiltonian
%
%
%%%%%%%%%%%%%%%
\be
H=H_0+V(x,y),
\ee
%%%%%%%%%%%%%%%
%
%
and we assume that the magnitude of the potential is small compared to the Landau level spacing $\omega_{\mathrm c}\gg |V|$. In this regime, we can neglect the coupling between different Landau levels and use degenerate perturbation theory to compute the degeneracy splitting of a single level.  Without loss of generality, we assume $V(x,y)\geqslant 0$ and $B>0$ in the following. 

%
%
%%%%%%%%%%%%%%%%%%%%%%%%%%%%%%%%%
%%%%%%%%%%%%%%%%%%%%%%%%%%%%%%%%%
\section{Variance of the LLL}
%%%%%%%%%%%%%%%%%%%%%%%%%%%%%%%%%
%%%%%%%%%%%%%%%%%%%%%%%%%%%%%%%%%
%
%
For simplicity, in the following, we focus on the LLL corresponding to $n=0$ for which the nonperturbed wavefunctions, in the thermodynamical limit,  can be chosen as
%
%
%%%%%%%%%%%%%%%
\be
\varphi_l(z)=\langle x,y|l\rangle=\frac{1}{\sqrt{2\pi \:l_B^2 \: l! \: 2^l}} \: z^l \: \e^{-|z|^2/4},
\label{eq:basis}
\ee
%%%%%%%%%%%%%%%
%
%
where $z=(x+iy)/l_B$, $l=0,1,...,N_\phi-1$ is the angular momentum, \mbox{$N_\phi=\mathcal{A}/(2\pi l_B^2)\gg 1$} is the degeneracy of the LLL  and  $l_B=1/\sqrt{B}$ is the magnetic length. 
%This set of eigenfunctions corresponds to the set given in Ref.~\onlinecite{Johnson49} [see Eq.~(45)].

To characterize the broadening of the LLL due to the potential, we consider its variance defined by
%
%
%%%%%%%%%%%%%%%
\be
w^2=\frac{1}{N_\phi} \sum_{p=0}^{N_\phi-1} \varepsilon_p^2 - \Bigg(\frac{1}{N_\phi} \sum_{p=0}^{N_\phi-1}\varepsilon_p \Bigg)^2,
\label{eq:def_variance}
\ee
%%%%%%%%%%%%%%%
%
%
where $\varepsilon_p$'s  are eigenenergies of $H$ projected onto the LLL. This variance can be recast as
%
%
%%%%%%%%%%%%%%%
\be
w^2=\frac{1}{N_\phi} \sum_{l=0}^{N_\phi-1}\sum_{l'=0}^{N_\phi-1}|\langle l | V |l'\rangle|^2 - \Bigg(\frac{1}{N_\phi} \sum_{l=0}^{N_\phi-1} \langle l | V |l\rangle\Bigg)^2,
\ee
%%%%%%%%%%%%%%%
%
%
so that one does not need to compute explicitly the $\varepsilon_p$'s.
Setting $\mathbf{r}=(x,y)=r(\cos \theta,\sin\theta)$, a matrix element  of the perturbation potential in the LLL basis \mbox{$\{|l\rangle, l=0,...,N_\phi-1\}$} reads
\beqn
\langle l| V|l' \rangle &=& \int \frac{\dd \mathbf{q}}{(2\pi)^2}\, \frac{\widetilde{V}(\mathbf{q})}{2\pi \sqrt{l! \, l' ! \, 2^{l+l'}}}   \\
&\times&\int_0^\infty \frac{\dd r}{l_B} \,\left(\frac{ r}{l_B}\right)^{1+l+l'} \e^{-\frac{r^2}{2 l_B^2}} \int_0^{2\pi} \dd \theta \:\e^{\im \mathbf{q}\cdot \mathbf{r}}\e^{\im \theta (l-l')},\nonumber
\eeqn
where we introduced the Fourier transform of the potential 
%
%
%%%%%%%%%%%%%%%
\be
\widetilde{V}(\mathbf{q})=\int \dd \mathbf{r}\, \e^{-\im \mathbf{q}\cdot \mathbf{r}} V(\mathbf{r}).
\ee
%%%%%%%%%%%%%%%
%
%
In the large-$N_\phi$ (thermodynamical) limit, one then gets:
\beqn
\sum_{l=0}^\infty \langle l| V|l \rangle &=& \frac{\widetilde{V}(0)}{2\pi l_B^2},\\
\sum_{l=0}^{\infty}\sum_{l'=0}^{\infty}|\langle l | V |l'\rangle|^2 &=& \frac{1}{2\pi l_B^2} \int \frac{\dd \mathbf{q}}{(2\pi)^2}\, |\widetilde{V}(\mathbf{q})|^2 \e^{-|\mathbf{q}|^2 l_B^2 /2}. \: \: \quad 
\eeqn
Finally, one obtains the following expression for the variance
%
%
%%%%%%%%%%%%%%%
\be
w^2=\int \frac{\dd \mathbf{q}}{(2\pi)^2}\, S(\mathbf{q}) \e^{- |\mathbf{q}|^2 l_B^2/2},
\label{LLLvar}
\ee
%%%%%%%%%%%%%%%
%
%
where we introduced the structure factor 
%
%
%%%%%%%%%%%%%%%
\be
S(\mathbf{q})=\frac{|\widetilde{V}(\mathbf{q})|^2}{\mathcal{A}}(1-\delta_{\mathbf{q},0}).
\label{eq:structure_factor_def}
\ee
%%%%%%%%%%%%%%%
%
%
Note that the term proportional to $\delta_{\mathbf{q},0}$ comes from the second term of Eq.~(\ref{eq:def_variance}) and is irrelevant only if $\widetilde{V}(\mathbf{0})=0$.
The variance is therefore essentially equal to the integral of the structure factor over a disk of radius $l_B^{-1}$ which is the main result of this paper. Before discussing the most interesting case of a potential with a dense Fourier spectrum, let us first show that this expression allows one to recover known results for simple potentials. 

%
%
%%%%%%%%%%%%%%%%%%%%%%%%%%%%%%%%%
%%%%%%%%%%%%%%%%%%%%%%%%%%%%%%%%%
\section{Periodic potential}
%%%%%%%%%%%%%%%%%%%%%%%%%%%%%%%%%
%%%%%%%%%%%%%%%%%%%%%%%%%%%%%%%%%
%
%
For a periodic potential of strength $V_0$ with a single spatial \mbox{frequency $a^{-1}$}
%
%
%%%%%%%%%%%%%%%
\be
V(x,y)=V_0 \left[ \cos(2 \pi x/a)+\cos(2 \pi y/a) \right],
\label{eq:periodic}
\ee
%%%%%%%%%%%%%%%
%
%
Eq.~(\ref{LLLvar}) leads to 
%
%
%%%%%%%%%%%%%%%
\be
w^2=V_0^2 \:\e^{-\frac{2 \pi^2}{B a^2}},
\label{eq:LLLvar_periodic}
\ee 
%%%%%%%%%%%%%%%
%
%
in agreement with the expression found by Rauh \cite{Rauh75} (see App.~\ref{app:var_periodic}  for  details). 
%Note that for periodic potentials, the natural set of eigenfunctions is rather the one given in Ref.~\onlinecite{Dana85_2} [see Eq.~(13)] but we can equally derive the above result with the basis given in Eq.~(\ref{eq:basis}).
The generalization to Fourier spectra  with a finite set of frequencies is straightforward, even if the potential is no longer periodic. In the zero-field limit, the LLL broadening is exponential and controlled by the smallest frequency. The case of a dense set of frequencies is more subtle. 

%
%
%%%%%%%%%%%%%%%%%%%%%%%%%%%%%%%%%
%%%%%%%%%%%%%%%%%%%%%%%%%%%%%%%%%
\section{ Random potential}
%%%%%%%%%%%%%%%%%%%%%%%%%%%%%%%%%
%%%%%%%%%%%%%%%%%%%%%%%%%%%%%%%%%
%
%
Landau level broadening due to an uncorrelated random potential has been widely studied in the literature \cite{Huckestein95}. 
For the simple case of a random potential  with zero mean and white-noise correlations
%
%
%%%%%%%%%%%%%%%
\beqn
\overline{V(\mathbf{r})}&=& 0, 
\label{eq:disorder_pot1}\\
\overline{V(\mathbf{r})V(\mathbf{r}')} &=& (V_0 \, a)^2 \delta(\mathbf{r}-\mathbf{r}'),
\label{eq:disorder_pot2}
\eeqn
%%%%%%%%%%%%%%%
%
%
where the overline denotes the average over disorder realizations,  Eq.~(\ref{LLLvar}) gives
%
%
%%%%%%%%%%%%%%%
\be
\overline{w^2}= V_0^2 \frac{a^2}{2\pi l_B^2} = V_0^2 \frac{B a^2}{2\pi},
\label{eq:LLLvar_disorder}
\ee
%%%%%%%%%%%%%%%
%
%
in agreement with the result of Ando \cite{Ando74_3} (see App.~\ref{app:var_random} for details). This result is very different from the one obtained for a potential  with a finite number of frequencies discussed above. 

For stealthy hyperuniform disorder \cite{Torquato18}, the structure factor is identically zero in a disk of radius $q_0>0$ around the origin. A reasonable approximation is to assume that $\overline{S(\mathbf{q}) }\propto \Theta(|\mathbf{q}|-q_0)$ leading to a LLL broadening
%
%
%%%%%%%%%%%%%%%
\be
\overline{w^2}\propto B \:\e^{-\frac{q_0^2}{2 B}},
\label{eq:LLLvar_stealthy}
\ee
%%%%%%%%%%%%%%%
%
%
intermediate between that of a periodic potential, Eq.~(\ref{eq:LLLvar_periodic}), and that of uncorrelated random disorder, Eq.~(\ref{eq:LLLvar_disorder}).

%
%
%%%%%%%%%%%
\begin{figure*}[t]
{\includegraphics[width=2.08 \columnwidth]{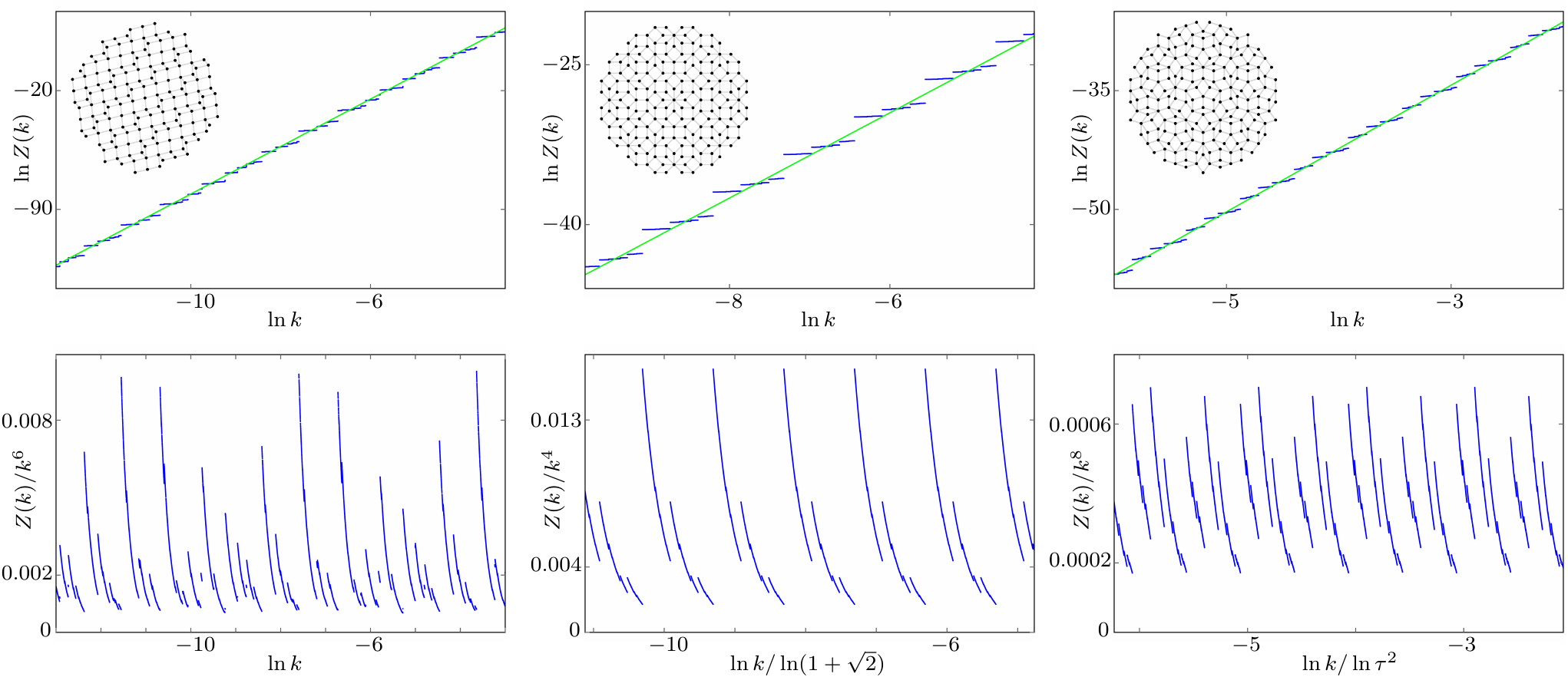}}
\caption{Integrated density function $Z$ of the Rauzy (left),  Ammann-Beenker (center), and Penrose  (right) tilings (see insets for illustrations). Top: Log-log plot (blue dots) together with the algebraic envelope $k^{2+\alpha}$ (green line). Bottom: $Z(k)/k^{2+\alpha}$ as a function of $\ln k/\ln \lambda$ ($\ln k$) showing 1-periodic (nonperiodic) oscillations for tilings with (without) discrete scale invariance.}
\label{fig}
\end{figure*}
%%%%%%%%%%%
%
%

%
%
%%%%%%%%%%%%%%%%%%%%%%%%%%%%%%%%%
%%%%%%%%%%%%%%%%%%%%%%%%%%%%%%%%%
\section{Potential with a dense Fourier spectrum}
%%%%%%%%%%%%%%%%%%%%%%%%%%%%%%%%%
%%%%%%%%%%%%%%%%%%%%%%%%%%%%%%%%%
%
%
The most interesting situation comes from potentials with a dense Fourier spectrum made of Bragg peaks (as found, e.g, in quasicrystals). To this aim,  let us consider a general potential 
%
%
%%%%%%%%%%%%%%%
\be
V(\mathbf{r})= V_0 \, a^2 \, \sum_{j=1}^N \delta(\mathbf{r}-\mathbf{r}_j),
\label{eq:potential_delta_def}
\ee
%%%%%%%%%%%%%%%
%
%
built on a set of $N$ scattering points located at position $\mathbf{r}_j$ with a typical density $a^{-2}$. The random potential discussed above belongs to this family. 

Before proceeding further, let us stress that the exponential term in Eq.~(\ref{LLLvar}) acts as a smooth cutoff that eliminates wave vectors $|\mathbf{q}| \gtrsim l_B^{-1}$. In order to analyze the behavior of $w^2$, we shall instead consider a sharp cutoff regularization by introducing the integrated intensity function 
%
%
%%%%%%%%%%%%%%%
\be
Z(k)=\int_{|\mathbf{q}|<k} \dd\mathbf{q}\,  S(\mathbf{q}),
\ee 
%%%%%%%%%%%%%%%
%
%
so that one has 
%
%
%%%%%%%%%%%%%%%
\be
w^2\sim Z(k\sim 1/l_B).
\label{eq:correspondance}
\ee 
%%%%%%%%%%%%%%%
%
%
This approximation clearly  misses exponentially small terms so that, for the periodic case discussed previously [see Eq.~(\ref{eq:LLLvar_periodic})], it  leads to $w^2=0$.  Let us remind that Eq.~(\ref{eq:correspondance}) is valid in the perturbative regime where $m V_0 \ll B$. In the following, we further focus on the case where $ B \ll 1/a^2$ since, for many potentials, $Z$ has a simple behavior in the  $k\sim 1/l_B\ll 1/a$ limit.

The  integrated intensity function (also known as the spectral measure \cite{Luck89}), is commonly used  to analyze sets of points with a nontrivial structure factor \cite{Oguz17}. 
In one-dimensional quasicrystals, $Z$ is conjectured to have a power-law envelope for $k\ll 1/a$ \cite{Luck89,Oguz17,Oguz19}. As we shall see, this is also the case for two-dimensional quasicrystals. Assuming \mbox{$Z(k) \underset{k \to 0}{\sim} k^{2+\alpha}$},  Eq.~(\ref{eq:correspondance}) leads to  
%
%
%%%%%%%%%%%%%%%
\be
w^2 \sim B^{\frac{2+\alpha}{2}},  \textrm{ for }m V_0 \ll B \ll1 /a^2,
\label{eq:main_result}
\ee 
%%%%%%%%%%%%%%%
%
%
establishing a relation between the broadening of the LLL  and the so-called hyperuniformity exponent $\alpha$ that characterizes the potential. For $\alpha>0$, the potential is hyperuniform \cite{Oguz17} whereas $\alpha<0$ refers to hypo-uniformity (or  anti-hyperuniformity \cite{Oguz19}). The special case $\alpha=0$ corresponds to a potential with a constant $S$, such as the random potential considered previously [see Eq.~(\ref{eq:LLLvar_disorder})].

Interestingly, if $Z$ further manifests a discrete scale invariance, i.e., if there exists $\lambda>1$ such that
%
%
%%%%%%%%%%%%%%%
\be
Z(k/\lambda)=Z(k)/\lambda^{2+\alpha},
\label{eq:lambda_def}
\ee
%%%%%%%%%%%%%%%
%
%
then one has
%
%
%%%%%%%%%%%%%%%
\be
Z(k)=k^{2+\alpha} F(\ln k/\ln\lambda),
\label{eq:dsi}
\ee
%%%%%%%%%%%%%%%
%
%
where $F(x+1)=F(x)$ (see the examples below and Refs.~\onlinecite{Sornette98,Akkermans13} for a review). As a result, the LLL variance $w^2$  displays log-periodic oscillations together with a power-law envelope in the small-$B$ limit. 

%
%
%%%%%%%%%%%%%%%%%%%%%%%%%%%%%%%%%
%%%%%%%%%%%%%%%%%%%%%%%%%%%%%%%%%
\section{Examples of quasiperiodic potentials}
%%%%%%%%%%%%%%%%%%%%%%%%%%%%%%%%%
%%%%%%%%%%%%%%%%%%%%%%%%%%%%%%%%%
%
%
For illustration, let us consider some potentials of the form given in Eq.~(\ref{eq:potential_delta_def}) where the points correspond to vertices of two-dimensional quasiperiodic tilings (see App.~\ref{app:Fourier}).  For each tiling considered below, we computed exactly the structure factor $S$, the hyperuniformity exponent $\alpha$ characterizing the power-law behavior of \mbox{$Z(k)\underset{k \to 0}{\sim} k^{2+\alpha}$}, and the discrete scale invariance factor $\lambda$ defined in Eq.~(\ref{eq:lambda_def}) when it exists. Numerical results displayed in Fig.~\ref{fig}  have been obtained by integrating more and more Bragg peaks of smaller and smaller intensities. In each case, we checked that the results were converged in the range considered. Units are taken such that $V_0=\sqrt{\mathcal{A}/(Na^2)}$ and  $a=1$, where $a$ is the edge length of the hypercubic lattice that is used to build the tiling in the standard cut-and-project method~\cite{DeBruijn81,Duneau85,Kalugin85,Elser85}.

Let us first consider the twofold-symmetric Rauzy tiling \cite{Vidal01}. The hyperuniformity exponent is $\alpha=4$ but $Z$  has no discrete scale invariance (see App.~\ref{app:Rauzy}). By contrast, for the eightfold-symmetric Ammann-Beenker \mbox{tiling \cite{Beenker82,Grunbaum87,Duneau89}}, the hyperuniformity exponent is \mbox{$\alpha=2$} and $Z$ has a discrete scale invariance with \mbox{$\lambda=1+\sqrt{2}$} (see App.~\ref{app:octo}).  For the fivefold-symmetric Penrose tilings, the hyperuniformity exponent is \mbox{$\alpha=6$}, and $Z$ has a discrete scale invariance with $\lambda=\tau^2$, where $\tau=\frac{1+\sqrt{5}}{2}$ is the golden ratio (see App.~\ref{app:Penrose}).

%
%
%%%%%%%%%%%%%%%%%%%%%%%%%%%%%%%%%
%%%%%%%%%%%%%%%%%%%%%%%%%%%%%%%%%
\section{Substitution tilings and discrete scale invariance}
%%%%%%%%%%%%%%%%%%%%%%%%%%%%%%%%%
%%%%%%%%%%%%%%%%%%%%%%%%%%%%%%%%%
%
%
As recently conjectured by O\u{g}uz et al. \cite{Oguz19}, the behavior of $Z$ in one-dimensional substitution tilings is determined by the eigenvalues of the substitution matrix. More precisely, for non-periodic binary substitutions associated with a  $2 \times 2$ substitution matrix with eigenvalues $\lambda_1>|\lambda_2| >0 $, one has 
 \be
Z\left(k/\lambda_1\right)=Z(k)(\lambda_2/\lambda_1)^2,
\ee
when $k$ tends to zero, so that 
\be
Z(k)= k^{1+\alpha}F(\ln k/\ln \lambda_1), \quad \alpha=1-2 \frac{\ln |\lambda_2|}{\ln \lambda_1},
\label{eq:zk1}
\ee 
with $F(x+1)=F(x)$. In two dimensions, it is very likely that the existence of substitution rules  with inflation/deflation also implies the existence of discrete scale invariance for $Z$. This is clearly the case for the  Ammann-Beenker and Penrose tilings, which, contrary to the Rauzy tiling, can be built by  inflation/deflation. However, we have not found a simple expression for the hyperuniformity exponent [such as the one given Eq.~(\ref{eq:zk1})] for two-dimensional binary substitutions.

%
%
%%%%%%%%%%%%%%%%%%%%%%%%%%%%%%%%%
%%%%%%%%%%%%%%%%%%%%%%%%%%%%%%%%%
\section{Outlook}
%%%%%%%%%%%%%%%%%%%%%%%%%%%%%%%%%
%%%%%%%%%%%%%%%%%%%%%%%%%%%%%%%%%
%
%
In this paper, we obtained a simple relation between the Landau level broadening and the integrated intensity function $Z$ of the pertur\-bing potential. For potentials with a dense Fourier spectrum made of Bragg peaks, this relation implies that the variance of the LLL is driven by the hyperuniformity exponent $\alpha$ [see Eq.~(\ref{eq:main_result})].  In the absence of a complete classification of the possible behavior of $Z$, a first step to go beyond would consist in analyzing two-dimensional potentials with a singular continuous Fourier spectrum for which one expects more complex behavior of $Z$ as observed in one dimension \cite{Luck89,Oguz19}. For instance, one may find noninteger exponents $\alpha$ or even nonalgebraic decay. Another important issue would be to consider the influence of Landau level mixing which is known to have dramatic effects on the localization properties of the eigenstates \cite{Haldane97}, and hence, on integer quantum Hall physics (see, e.g., Ref.~\onlinecite{Kivelson92}). 
 It would also be important to bridge the gap between the perturbed free-electron results and the one obtained numerically in tight-binding models \cite{Fuchs18}. A possible route would be to develop the analog of Wilkinson semi-classical treatment \cite{Wilkinson84} for nonperiodic potential. 

Finally, let us mention that the magnetic-field dependence of the Landau level broadening induced by disorder has already been  measured in graphene \cite{Orlita08}. Combining such an experimental device with a nontrivial superlattice potential would allow us to measure the behaviors discussed in the present work.

%%%%%%%%%%%%%%%%%%%%%%%%%%%%%%%%%
%%%%%%%%%%%%%%%%%%%%%%%%%%%%%%%%%
\acknowledgments

We thank M. Duneau, T. Fernique, and J.-M. Luck for fruitful discussions.

%%%%%%%%%%%%%%%%%%%%%%%%%%%%%%%%%
%%%%%%%%%%%%%%%%%%%%%%%%%%%%%%%%%
%%%%%%%%%%%%%%%%%%%%%%%%%%%%%%%%%
%%%%%%%%%%%%%%%%%%%%%%%%%%%%%%%%%
%%%%%%%%%%%%%%%%%%%%%%%%%%%%%%%%%

\appendix

\begin{widetext}

%
%
%%%%%%%%%%%%%%%%%%
\section{Variance for a periodic potential}
\label{app:var_periodic}
%%%%%%%%%%%%%%%%%%
%
%

The Fourier transform of the periodic potential (\ref{eq:periodic}) is given by
\beqn
\widetilde{V}(\mathbf{q})&=&\frac{V_0}{2} (2\pi)^2 \left[ \delta(q_x-2\pi/a) \delta(q_y)+ \delta(q_x+2\pi/a) \delta(q_y)+ \delta(q_x) \delta(q_y-2\pi/a)+ \delta(q_x) \delta(q_y+2\pi/a) \right] , \\
&=&\frac{V_0}{2} \mathcal{A} \left[ \delta_{q_x,2\pi/a} \delta_{q_y,0}+  \delta_{q_x,-2\pi/a} \delta_{q_y,0}+ \delta_{q_x,0} \delta_{q_y,2\pi/a}+ \delta_{q_x,0} \delta_{q_y,-2\pi/a} \right], 
\eeqn
where we used the fact that $\mathcal{A} \delta_{\mathbf{q},0}=(2\pi)^2 \delta(\mathbf{q})$, in the thermodynamic limit. The structure factor (\ref{eq:structure_factor_def}) becomes
\beqn
S(\mathbf{q})&=&\frac{(1-\delta_{\mathbf{q},0})}{\mathcal{A}}\left| \frac{V_0}{2} \mathcal{A} \left[ \delta_{q_x,2\pi/a} \delta_{q_y,0}+  \delta_{q_x,-2\pi/a} \delta_{q_y,0}+ \delta_{q_x,0} \delta_{q_y,2\pi/a}+ \delta_{q_x,0} \delta_{q_y,-2\pi/a} \right] \right|^2,   \\
&=&\frac{V_0^2}{4} \mathcal{A} \left[ \delta_{q_x,2\pi/a} \delta_{q_y,0}+  \delta_{q_x,-2\pi/a} \delta_{q_y,0}+ \delta_{q_x,0} \delta_{q_y,2\pi/a}+ \delta_{q_x,0} \delta_{q_y,-2\pi/a} \right] , \\
&=&\frac{V_0^2}{4} (2\pi)^2 \left[ \delta(q_x-2\pi/a) \delta(q_y)+ \delta(q_x+2\pi/a) \delta(q_y)+ \delta(q_x) \delta(q_y-2\pi/a)+ \delta(q_x) \delta(q_y+2\pi/a) \right], 
\eeqn
where we used the Kronecker delta in order to compute the modulus square of the Fourier transform (otherwise, the square of a Dirac delta is ill-defined). Then Eq.~(\ref{LLLvar}) involves the integral of Dirac-delta functions, which straightforwardly leads to Eq.~(\ref{eq:LLLvar_periodic}).

%
%
%%%%%%%%%%%%%%%%%%
\section{Variance for a random potential}
\label{app:var_random}
%%%%%%%%%%%%%%%%%%
%
%

The Fourier transform of the uncorrelated random potential defined by Eqs.~(\ref{eq:disorder_pot1})-(\ref{eq:disorder_pot2}) is given by 
\beqn
\overline{\widetilde{V}(\mathbf{q})} = \int \text{d} \mathbf{r} \text{e}^{- \text{i} \mathbf{r}\cdot \mathbf{q}} \overline{V(\mathbf{r})}=0, \nonumber
\eeqn
and
\beqn
\overline{|\widetilde{V}(\mathbf{q})|^2} = \int \text{d} \mathbf{r} \text{e}^{- \text{i} \mathbf{r}\cdot \mathbf{q}}\int \text{d} \mathbf{r}' \text{e}^{\text{i} \mathbf{r}'\cdot \mathbf{q}} \overline{V(\mathbf{r})V(\mathbf{r}')}
=(V_0 a)^2\mathcal{A}. \nonumber
\eeqn

The structure factor (\ref{eq:structure_factor_def}) becomes $\overline{S(\mathbf{q})}= (V_0 a)^2 (1-\delta_{\mathbf{q},0})$.
In Eq.~(\ref{LLLvar}) the Kronecker delta does not contribute and the Gaussian integral gives Eq.~(\ref{eq:LLLvar_disorder}):
\beqn
\overline{w^2}&=& \frac{(V_0 a)^2}{(2\pi)^2} \int \text{d} \mathbf{q} \: \text{e}^{-|\mathbf{q}|^2 l_B^2/2}=\frac{(V_0 a)^2}{(2\pi)^2} \frac{2\pi}{l_B^2}. \nonumber
\eeqn

\end{widetext}

 %
%%%%%%%%%%%%%%%%%%%%%%%%%%%%%%%%%
%%%%%%%%%%%%%%%%%%%%%%%%%%%%%%%%%
\section{Fourier transform of a cut-and-project quasicrystal}
\label{app:Fourier}
%%%%%%%%%%%%%%%%%%%%%%%%%%%%%%%%%
%%%%%%%%%%%%%%%%%%%%%%%%%%%%%%%%%
%

The cut-and-project (CP) method \cite{DeBruijn81,Duneau85,Kalugin85,Elser85} consists in selecting points of a $D$-dimensional periodic lattice if their projection onto the $(D-d)$-dimensional perpendicular space $E_\perp$ belongs to a so-called acceptance window. The tiling is then obtained by projecting these selected points onto the complementary $d$-dimensional parallel space $E_\varparallel$.

Any vector $\mathbf{v}$ in hyperspace can be decomposed uniquely in terms of its projection onto parallel and perpendicular spaces as
\be
\mathbf{v}=\mathbf{v}_\varparallel + \mathbf{v}_\perp.
\ee

As explained in the early papers introducing the CP method \cite{Duneau85,Kalugin85,Elser85}, the Fourier transform of quasiperiodic tilings can be computed from the higher-dimensional space it stems from. The main idea is that since points of the tiling are selected from a periodic tiling via an acceptance window, computing the Fourier transform of the tiling essentially amounts to compute the Fourier transform of this acceptance window. 

For a tiling with $N$ sites (vertices) at position $\mathbf{R}_j^\varparallel$ and obtained by the CP method, the microscopic density is
%
%
%%%%%%%%%%%%%%
\be 
n({\bf r}_\varparallel)=\sum_{j =1}^N \delta({\bf r}_\varparallel-{{\bf R}_j^\varparallel}),
\ee
%%%%%%%%%%%%%%
%
%
and its Fourier transform is
%
%
%%%%%%%%%%%%%%
\beq
\widetilde{n}({\bf q}_\varparallel) = \sum_{j=1}^N {\rm e}^{-{\rm i} \, {\bf q}_\varparallel.{{\bf R}_j^\varparallel}},
\label{eq:nt}
\eeq
%%%%%%%%%%%%%%
%
%
where the sum runs over all sites of the $d$-dimensional tiling considered. The convention that we use is that capital letters (such as ${\bf R}_j^\varparallel$) refer to discrete points and small letters (such as ${\bf r}_\varparallel$) to a continuum of points.

Let ${\bf R}$ be a point of the $D$-dimensional hypercubic lattice and ${\bf K}$ a vector of its reciprocal lattice such that ${\bf K}\cdot {\bf R}=2\pi \times$integer. These vectors can be decomposed onto the parallel and perpendicular spaces such that their scalar product reads
%
%
%%%%%%%%%%%%%%
\be
\mathbf{K}\cdot \mathbf{R} = \mathbf{K}_\varparallel \cdot \mathbf{R}_\varparallel +\mathbf{K}_\perp \cdot \mathbf{R}_\perp =2\pi \times \textrm{integer}. 
\label{eq:kr}
\ee
%%%%%%%%%%%%%%
%
%
Equation~(\ref{eq:nt}) is non-zero iff ${\bf q}_\varparallel = \mathbf{K}_\varparallel$, in which case it becomes
%
%
%%%%%%%%%%%%%%
\beq
\widetilde{n}(\mathbf{K}_\varparallel)=\sum_{j=1}^N {\rm e}^{{\rm i} \, {\bf K}_\perp.{{\bf R}_j^\perp}}.
\eeq
%%%%%%%%%%%%%%
%
%

For a quasicrystal built along an irrational plane (parallel space), the points in perpendicular space densely and uniformly fill the acceptance window such that
%
%
%%%%%%%%%%%%%%
\beq
\widetilde{n}({\bf K}_\varparallel)=N \int_{\mathcal{A}_\perp} \frac{\text{d}\mathbf{r}_\perp}{\mathcal{A}_\perp} {\rm e}^{{\rm i} \, {\bf K}_\perp.{{\bf r}_\perp}},
\label{eq:ntf2}
\eeq
%%%%%%%%%%%%%%
%
%
where the integral is over the acceptance window in perpendicular space and $\mathcal{A}_\perp$ is its $(D-d)$-dimensional volume. 

Now, for any vector ${\bf q}_\varparallel$ in parallel space, the Fourier transform of the density Eq.~(\ref{eq:nt}) reads
%
%
%%%%%%%%%%%%%%
\beq 
\widetilde{n}({\bf q}_\varparallel)=\sum_{\mathbf{K}} \delta_{\mathbf{q}_\varparallel,\mathbf{K}_\varparallel} N \int_{\mathcal{A}_\perp} \frac{\text{d}\mathbf{r}_\perp}{\mathcal{A}_\perp} {\rm e}^{{\rm i} \, {\bf K}_\perp.{{\bf r}_\perp}}, 
\label{eq:ntf}
\eeq
%%%%%%%%%%%%%%
%
%
where the sum is performed over all vectors $\mathbf{K}$ of the reciprocal lattice of the hypercubic lattice. 
As we are considering a quasicrystal, for any $\mathbf{K}_\varparallel$ there is a unique $\mathbf{K}$ and therefore $\mathbf{K}_\perp$ is well defined. 
If $\{\mathbf{a}_j^*;\, j=1,...,D\}$ is a basis of vectors in reciprocal space, then $\mathbf{K}=\sum_j n_j \mathbf{a}_j^*$, where $n_j$ are integers. Its parallel and perpendicular components are also functions of the same integers:
%
%
%%%%%%%%%%%%%%
\beq 
\mathbf{K}=\mathbf{K}_\varparallel (n_1,...,n_D)+\mathbf{K}_\perp (n_1,...,n_D).
\eeq
%%%%%%%%%%%%%%
%
%
Therefore, the sum over $\mathbf{K}$ in Eq.~(\ref{eq:ntf}) is actually a sum over $D$ integers $n_1,...,n_D$, clearly showing that the Fourier transform is pure point  of rank $D> d$.

Let us define the structure factor in the thermodynamical limit as
%
%
%%%%%%%%%%%%%%
\be
S({\bf q}_\varparallel)=\frac{|\widetilde{n}({\bf q}_\varparallel)|^2}{N} \left(1- \delta_{\bf{q_\varparallel,0}} \right).
\label{eq:sf}
\ee
%%%%%%%%%%%%%%
%
%

For two-dimensional potentials ($d=2$) of the form given by Eq.~(\ref{eq:potential_delta_def}), this definition differs from the one given in Eq.~(\ref{eq:structure_factor_def}) by a  factor $V_0^2 a^4 N/\mathcal{A}$, which disappears upon choosing units such that \mbox{$V_0=\sqrt{\mathcal{A}/(Na^2)}$}.

As explained in Ref.~\onlinecite{Oguz17},  for a spectrum made of a dense set  of Bragg peaks (discontinuous $S$), the integrated intensity function 
%
%
%%%%%%%%%%%%%%
\beq
Z(k)=\int_{|{\bf q}_\varparallel|<k} S({\bf q}_\varparallel) \, {\rm d} {\bf q}_\varparallel,
\eeq
%%%%%%%%%%%%%%
%
%
provides a reliable characterization of the point distribution. Here, the integral is performed over a disk of radius $k$. This function is also known as the spectral measure in Ref~\cite{Luck89}. 

For $d$-dimensional tilings built by the CP method, this quantity can be recast in the following form:
%
%
%%%%%%%%%%%%%%
\beq
Z(k)=\frac{(2\pi)^d}{\mathcal{A}} \sum_{|{\bf K}_\varparallel | <k} S({\bf K}_\varparallel). 
\label{eq:Zk}
\eeq
%%%%%%%%%%%%%%
%
%

%
%%%%%%%%%%%%%%%%%%%%%%%%%%%%%%%%%
%%%%%%%%%%%%%%%%%%%%%%%%%%%%%%%%%
\section{The Rauzy tiling}
\label{app:Rauzy}
%%%%%%%%%%%%%%%%%%%%%%%%%%%%%%%%%
%%%%%%%%%%%%%%%%%%%%%%%%%%%%%%%%%
%

%
%
%%%%%%%%%%%%%%%%%%
\subsection{Fourier Transform}
%%%%%%%%%%%%%%%%%%
%
%

The two-dimensional (generalized) Rauzy tiling has been introduced in  Ref.~\onlinecite{Vidal01}. This is a codimension-one tiling built from the cubic lattice $\mathbb{Z}^3$ (edge length $a=1$) with a one-dimensional perpendicular space oriented along the direction  ${\bf e}_\perp=(\theta^2,\theta,1)$ where $\theta$ is the real (Pisot-Vijayaraghavan) root of the cubic equation \mbox{$x^3=x^2+x+1$}. Contrary to the Ammann-Beenker and the Penrose tilings discussed in the next sections, the Rauzy tiling cannot be built by substitution rules. 

For such a codimension-one quasicrystal, the acceptance window is a segment of length $\mathcal{A}_\perp$ defined as the projection of  $\mathbf{h}=(1,1,1)$ onto the perpendicular space. This acceptance window also corresponds to the projection of the unit cube onto the perpendicular space.
In this case, Eq.~(\ref{eq:ntf2}) gives:
%
%
%%%%%%%%%%%%%%
\be
|\widetilde{n}({\bf K}_\varparallel)|=N \left| \text{sinc} \left(\frac{{\bf K}_\perp.{\bf h}_\perp}{2}\right) \right|.
\label{eq:TF_Rauzy_fin}
\ee
%%%%%%%%%%%%%%
%
%

%
%
%%%%%%%%%%%%%%%%%%
\subsection{Structure factor}
%%%%%%%%%%%%%%%%%%
%
%

The structure factor is defined in Eq.~(\ref{eq:sf}). Our goal is to analyze the behavior of  $S(\mathbf{q}_\varparallel)$ in the limit where $|\mathbf{q}_\varparallel|$ tends to zero. By definition, one has $S(0)=0$ but its behavior for small $|\mathbf{q}_\varparallel|$ is nontrivial since $S(\mathbf{q}_\varparallel) \neq 0$ only when $\mathbf{q}_\varparallel$ coincides with the parallel component $\mathbf{ K}_\varparallel$ of a reciprocal-lattice vector ${\bf K}$ of the cubic lattice. Thus, we are interested in computing the behavior of $S$ when $|{\bf K}_\varparallel|$ goes to $0$ for  ${\bf K}_\varparallel \neq 0$:

%
%
%%%%%%%%%%%%%%
\beqn
S({\bf K}_\varparallel)&=& N \left|\frac {\sin\Big(\frac{{\bf K}_\perp.{\bf h}_\perp}{2}\Big)}{\frac{{\bf K}_\perp.{\bf h}_\perp}{2}} \right|^2,\\
&=& N \left|\frac {\sin\Big(\frac{{\bf K}_\varparallel.{\bf h}_\varparallel}{2}\Big)}{\frac{{\bf K}_\perp.{\bf h}_\perp}{2}} \right|^2,\\
&\underset{|{\bf K}_\varparallel| \to 0}{\simeq}& N \left| \frac{{\bf K}_\varparallel.{\bf h}_\varparallel}{{\bf K}_\perp.{\bf h}_\perp} \right|^2,\\ \nonumber
\label{eq:struc_fact_def}
\eeqn
%%%%%%%%%%%%%%
%
%
where we used the fact that ${\bf K}$ is a reciprocal-lattice vector and ${\bf h}$ a direct-lattice vector. The difficulty comes from the fact that, when $|{\bf K}_\varparallel |$ goes to $0$, $|{\bf K}_\perp |$ diverges. So, the goal is to find the relation between these two components.

One way to investigate this issue is to follow the approach proposed in Ref.~\onlinecite{Oguz17} for the Fibonacci chain (see App.~\ref{app:Fibonacci}).  In the $\mathbb{Z}^3$ canonical basis, any reciprocal-lattice vector ${\bf K}$ has coordinates $2\pi (l,m,n)$ where $l, m$, and $n$ are integers. To analyze the behavior of ${\bf K}_\varparallel.{\bf h}_\varparallel$ and ${\bf K}_\perp.{\bf h}_\perp$, let us consider the matrix 
%
%
%%%%%%%%%%%%%%%%%%%%%%%%%%%%%%%%%
\begin{eqnarray}
M=\left( 
\begin{array}{ccc}
                      1 & 1 & 1 \\
                      1 & 0 & 0 \\
                      0 & 1 & 0 
\end{array}
         \right)
\mbox{ , }
\label{eq:mrauzy}
\end{eqnarray}
%%%%%%%%%%%%%%%%%%%%%%%%%%%%%%%%%
%
%
that satisfies $M^{3}=M^{2}+M+1$. The eigenvalues of $M$ are the Tribonacci constant $\theta\simeq 1.8393 $ and two complex conjugate eigenvalues $\rm{e}^{\pm \rm{i} \phi}/ \sqrt{\theta}$ with $\phi\simeq 2.1762$. The eigenvector associated to $\theta$ corresponds to the perpendicular direction ${\bf e}_\perp$.  Since $\theta$ is a Pisot-Vijayaraghavan number, the action of $M^p$ onto any vector ${\bf v}$, such that ${\bf v}.{\bf e}_\perp \neq 0$, drives this vector towards the direction ${\bf e}_\perp$ in the large-$p$ limit. 

Hence, to analyze the behavior of $S$ in the limit where ${\bf K}_\varparallel$ tends to zero  [see Eq.~(\ref{eq:struc_fact_def})], let us consider \mbox{${\bf K}^{(p)}=M^p {\bf K}$}. More precisely, we are interested in computing  ${\bf K}_\varparallel^{(p)}.{\bf h}_\varparallel$ and ${\bf K}^{(p)}_\perp.{\bf h}_\perp$. Keeping in mind that ${\bf h}_\perp=P_\perp (1,1,1)$ and ${\bf h}_\varparallel=(\mathds{1}-P_\perp) (1,1,1)$ (where $P_\perp$ is the projector onto the perpendicular space), one can easily compute these quantities. After some algebra, one gets:
\begin{widetext}
%
%
%%%%%%%%%%%%%%
\beqn
{\bf K}_\varparallel^{(p)}.{\bf h}_\varparallel&=&\frac{(\theta -1) \theta ^{-\frac{p+1}{2}} \left\{\sin (p \, \phi ) \left[(1+\theta ^2)  m-\theta  (l+n)\right]+\sqrt{\theta } \left[(\theta  m-l) \sin ((1+p)
   \phi)+(\theta  n-m) \sin ((1-p)
   \phi)\right]\right\}}{ \sin(\phi )[(\theta -1) \theta +1]} ,\nonumber \\   
   &&  \label{eq:gpara_Rauzy}\\
  {\bf K}_\perp^{(p)}.{\bf h}_\perp&=& \frac{1}{\sin(\phi )[(\theta -1) \theta +1] \left(2 \theta ^{3/2} \cos (\phi )-\theta ^3-1\right)}  \times 
 \Big\{-\theta ^p \left(\theta ^4+\theta ^2+1\right) \sin (\phi ) \left(\theta  l-2 \sqrt{\theta } m \cos (\phi )+n\right) \nonumber \\
 &&+\theta ^{-\frac{p-1}{2}} \Big[\theta ^{3/2} \sin ((p+1) \phi) \big[(\theta -1) l+\theta ^2 (n-m)\big] +(\theta -1) \theta  \sin (p \, \phi ) \big[l-(\theta +1) m+\theta  n\big] \nonumber \\
&& +\sqrt{\theta } \sin ((p-1) \phi ) [-l+m+(\theta -1) \theta  n]+\theta ^3 (l-\theta  m) \sin ((p+2) \phi )+ (m-\theta  n) \sin ((p-2) \phi ) \Big] \Big\}.\label{eq:gperp_Rauzy}
\eeqn
%%%%%%%%%%%%%%
%
%
\end{widetext}

Thus, in the large-$p$ limit,  one finds that ${\bf K}_\varparallel^{(p)}.{\bf h}_\varparallel$ vanishes as $\theta^{-p/2}$ ,  ${\bf K}_\perp^{(p)}.{\bf h}_\perp$ diverges as $\theta^{p}$, and $S({\bf K}_\varparallel^{(p)})$ behaves as $\theta^{-3p}$.
As a result, one finds that 
%
%
%%%%%%%%%%%%%%
\beq
S({\bf K}_\varparallel) \underset{|{\bf K}_\varparallel| \to 0}{\sim} |{\bf K}_\varparallel|^6,
\eeq
%%%%%%%%%%%%%%
%
%
for all $(l,m,n)$. However, we emphasize that, contrary to the Fibonacci chain Ref.~\onlinecite{Oguz17} (see also App.~\ref{app:Fibonacci}), this  power-law behavior is modulated by a bounded oscillating nonperiodic function as can be seen in Eqs.~(\ref{eq:gpara_Rauzy}-\ref{eq:gperp_Rauzy}). 

%
%
%%%%%%%%%%%%%%%%%%
\subsection{Integrated intensity function}
%%%%%%%%%%%%%%%%%%
%
%

The integrated intensity function is defined in Eq.~(\ref{eq:Zk}). The sum over all vector ${\bf K}_\varparallel$ with a norm smaller than $k$ can be decomposed into a sum over all triplets $(l,m,n)$ and their iterated under $M^p$. 
As a result, one has:
%
%
%%%%%%%%%%%%%%
\beq
Z(k)=\frac{4\pi^2}{\mathcal A} \sum_{(l,m,n)} \sum_{p=p_{(l,m,n)}}^\infty S({\bf K}_\varparallel^{(p)}), 
\label{eq:Z}
\eeq
%%%%%%%%%%%%%%
%
%
where $p_{(l,m,n)}$ is the smallest integer fulfilling the constraint \mbox{$|{\bf K}_\varparallel^{(p)}|<k$}. As already discussed in the previous section, in the large-$p$ limit, 
%
%
%%%%%%%%%%%%%%
\beqn
S({\bf K}_\varparallel^{(p)}) &\simeq&  |{\bf K}_\varparallel^{(p)}|^6 f(l,m,n,p), \label{eq:scaling_S}\\
|{\bf K}_\varparallel^{(p)}| &\simeq&  \theta^{-p/2} g(l,m,n,p), \label{eq:scaling_K}\\
&& \nonumber
\eeqn
%%%%%%%%%%%%%%
%
%
where, for a given triplet $(l,m,n)$, $f$ and $g$ are bounded oscillating function of $p$  [see Eqs.~(\ref{eq:gpara_Rauzy})-(\ref{eq:gperp_Rauzy})]).  Thus, $S$ is bounded both above and below
%
%
%%%%%%%%%%%%%%
\be
Z^-(k) \leqslant Z(k) \leqslant Z^+(k),
\ee
%%%%%%%%%%%%%%
%
%
 where 
%
%
%%%%%%%%%%%%%%
\beqn
Z^\pm(k)&=&\frac{4\pi^2}{\mathcal A} \sum_{(l,m,n)} c^\pm(l,m,n) \sum_{p=p_{(l,m,n)}}^\infty \theta^{-3 p}, \quad \quad  \quad\label{eq:def_Z}\\
c^+(l,m,n)&=&\max_p  f(l,m,n,p)^6 g(l,m,n,p),\\
c^-(l,m,n)&=&\min_p    f(l,m,n,p)^6 g(l,m,n,p).
\eeqn
%%%%%%%%%%%%%%
%
%
Interestingly,  $Z^\pm(k/\sqrt{\theta})=Z^\pm (k)/\theta^{3} $, as can be seen from Eqs.~(\ref{eq:scaling_S})-(\ref{eq:scaling_K}) since dividing $k$ by $\sqrt{\theta}$ simply amounts to change $p_{(l,m,n)}$ into $p_{(l,m,n)}+1$ in Eq.~(\ref{eq:def_Z}). Such a relation reflects a discrete scale invariance \cite{Sornette98} (see also next section) for $Z^\pm$ and implies a power-law envelope 
%
%
%%%%%%%%%%%%%%
\be 
Z(k) \underset{k \to 0}{\sim} k^6.
\ee
%%%%%%%%%%%%%%
%
%
Note that, despite the fact that $Z$ is defined as an integral of $S$, they are both characterized by a power law with the same exponent. This is a consequence of the fact that $S$ is discontinuous (discrete) and dense.
 
%
%%%%%%%%%%%%%%%%%%%%%%%%%%%%%%%%%
%%%%%%%%%%%%%%%%%%%%%%%%%%%%%%%%%
\section{Integrated intensity function of the Fibonacci chain}
\label{app:Fibonacci}
%%%%%%%%%%%%%%%%%%%%%%%%%%%%%%%%%
%%%%%%%%%%%%%%%%%%%%%%%%%%%%%%%%%
%
%

%
%%%%%%%%%%%
\begin{figure}
\includegraphics[width=0.9\columnwidth]{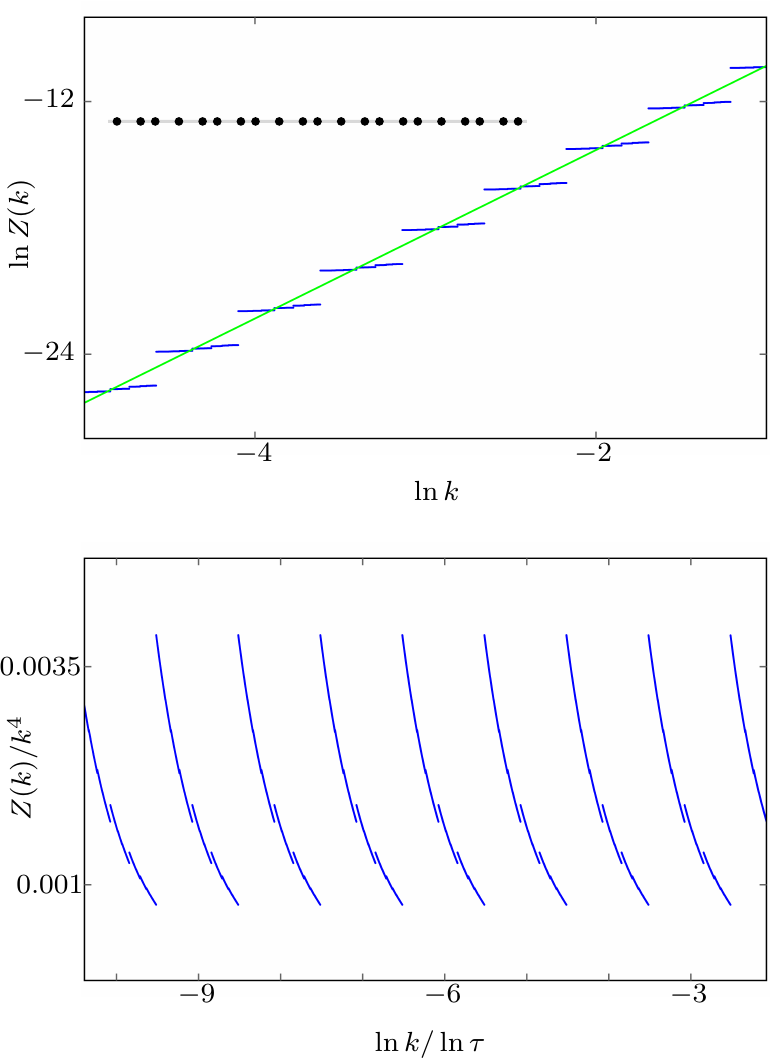}
\caption{Integrated density function $Z$ of the Fibonacci chain (see inset). Top: Log-log plot (blue dots) together with the power-law envelope $k^4$  (green line). Bottom: $Z(k)/k^4$ as a function of $\ln k/\ln \tau$ showing periodic oscillations (with period 1).}
\label{fig:Fibo}
\end{figure}
%%%%%%%%%%%
%
%
The Fibonacci chain is a one-dimensional tiling built from the square lattice $\mathbb{Z}^2$ (edge length $a=1$).
The integrated intensity function $Z$ of the Fibonacci chain has been widely discussed in Ref.~\onlinecite{Oguz17}. However, one important property has been missed. As a codimension-one system, the Fourier transform of the Fibonacci chain can be easily computed. The structure factor is
%
%
%%%%%%%%%%%%%%%%%%%%%%%%%%%%%%%%%
\be 
S({\bf K}_\varparallel)=N \left|\frac {\sin\Big(\frac{{\bf K}_\varparallel.{\bf h}_\varparallel}{2}\Big)}{\frac{{\bf K}_\perp.{\bf h}_\perp}{2}} \right|^2,
\ee
%%%%%%%%%%%%%%%%%%%%%%%%%%%%%%%%%
%
%
where ${\bf h}_\varparallel$ and ${\bf h}_\perp$ are the projections of the vector  \mbox{${\bf h}=(1,1)$} onto \mbox{${\bf e}_\varparallel=\frac{1}{\sqrt{1+\tau^2}}(-1,\tau)$} and \mbox{${\bf e}_\perp=\frac{1}{\sqrt{1+\tau^2}}(\tau,1)$}, where $\tau=\frac{1+\sqrt{5}}{2}$ is the golden ratio. The integrated intensity function is then 
%
%
%%%%%%%%%%%%%%%%%%%%%%%%%%%%%%%%%
\be
Z(k)=\frac{2\pi}{\mathcal{A}} \sum_{|{\bf K}_\varparallel| <k} S({\bf K}_\varparallel),
\label{eq:d2}
\ee
%%%%%%%%%%%%%%%%%%%%%%%%%%%%%%%%%
%
%
where  $\mathcal{A}$ is the total length of the chain.  As for the Rauzy tiling, let us consider the matrix 
%
%
%%%%%%%%%%%%%%%%%%%%%%%%%%%%%%%%%
\begin{eqnarray}
M=\left( 
\begin{array}{cc}
                      1 & 1  \\
                      1 & 0                 
\end{array}
         \right)
\mbox{ , }
\end{eqnarray}
%%%%%%%%%%%%%%%%%%%%%%%%%%%%%%%%%
%
%
that satisfies $M^{2}=M+1$. Eigenspaces of $M$ correspond to the perpendicular and parallel directions with eigenvalues $\tau$ and $-1/\tau$, respectively. The small-$k$ behavior of $Z$ is obtained by analyzing sequences \mbox{${\bf K}^{(p)}=M^p {\bf K}$} (see Ref.~\onlinecite{Oguz17}). One then gets, in the large-$p$ limit, 
%
%
%%%%%%%%%%%%%%
\beqn
S({\bf K}_\varparallel^{(p)}) &\simeq& |{\bf K}_{\varparallel}^{(p)}|^4 f(l,m), \label{eq:scaling_S_Fibo}\\
|{\bf K}_\varparallel^{(p)}| &\simeq&  \tau^{-p} g(l,m) \label{eq:scaling_K_Fibo}.
\eeqn
%%%%%%%%%%%%%%
%
%
However, contrary to the Rauzy tiling, $f$ and $g$ do not depend on $p$. Thus, following the same line of reasoning as above, one straightforwardly gets the discrete scaling relation
%
%
%%%%%%%%%%%%
\be
Z(k/\tau)=Z(k)/\tau^{4}.
\ee 
%%%%%%%%%%%%
%
% 
The solution of this equation can be written as
%
%
%%%%%%%%%%%%
\be
Z(k)=k^4 F(\ln k/\ln \tau),
\ee 
%%%%%%%%%%%%
%
% 
where $F(x+1)=F(x)$ (for a review on discrete scale invariance, see Ref.~\onlinecite{Sornette98}). As a result, $Z$ has a power-law envelope together with log-periodic oscillations (see Fig.~\ref{fig:Fibo} for illustration). This is in stark contrast with the  Rauzy tiling where only $Z^+$ and $Z^-$  obey such a discrete scale invariance but not $Z$ itself. Practically, to compute $Z$, we first select a set of $\mathbf{K}$ points in the reciprocal lattice of $\mathbb{Z}^2$ inside a given ball of radius $K_\text{max}$ around the origin. For each of these points, we consider the sequence of points $\mathbf{K}^{(p)}$ with $p=0,...,p_\text{max}$, and we compute $S$ for each corresponding $\mathbf{K}_\varparallel^{(p)}$ (avoiding possible redundancy). $Z$ is then obtained by summing over these Bragg peaks according to Eq.~(\ref{eq:d2}). We check the convergence of the results displayed in Fig.~\ref{fig:Fibo}  by increasing $K_\text{max}$ and $p_\text{max}$.

%
%%%%%%%%%%%%%%%%%%
%%%%%%%%%%%%%%%%%%
\section{The octagonal tiling}
\label{app:octo}
%%%%%%%%%%%%%%%%%%
%%%%%%%%%%%%%%%%%%
%

%
%
%%%%%%%%%%%%%%%%%%
\subsection{Fourier transform}
%%%%%%%%%%%%%%%%%%
%
%

The octagonal (Ammann-Beenker) tiling \cite{Beenker82,Grunbaum87,Duneau89} is a codimension-two tiling built from the four-dimensional  hypercubic lattice $\mathbb{Z}^4$ (edge length $a=1$). Perpendicular and  parallel spaces are spanned by the eigenvectors of the matrix
%
%
%%%%%%%%%%%%%%%%%%%%%%%%%%%%%%%%%
\begin{eqnarray}
M=\left( 
\begin{array}{cccc}
                      1 & 0 & -1 & 1\\
                      0 & 1 & -1 & -1\\
                      -1 & -1 & 1 & 0\\
                     1 & -1 & 0 & 1
\end{array}
         \right)
\mbox{ , }
\end{eqnarray}
%%%%%%%%%%%%%%%%%%%%%%%%%%%%%%%%%
%
%
associated to eigenvalues $\lambda_\pm=1\pm\sqrt{2}$.  This matrix satisfies $M^2=2M+1$ and its eigenvalues are twofold degenerate. Here, we choose the following orthonormal eigenbasis
%
%
%%%%%%%%%%%%%%
\beqn
 {\bf e}_{\varparallel,1}&=& \left(-\frac{1}{2},\frac{1}{2},0,\frac{1}{\sqrt{2}}\right), \nonumber \\
 {\bf e}_{\varparallel,2}&=&  \left(\frac{1}{2},\frac{1}{2},\frac{1}{\sqrt{2}},0\right), \nonumber \\
 {\bf e}_{\perp,1}&=& \left(\frac{1}{2},-\frac{1}{2},0,\frac{1}{\sqrt{2}}\right), \nonumber \\
 {\bf e}_{\perp,2}&=& \left(-\frac{1}{2},-\frac{1}{2},\frac{1}{\sqrt{2}},0\right),
\eeqn
%%%%%%%%%%%%%%
%
%
where the perpendicular (parallel) space is associated to $\lambda_+$ ($\lambda_-$). 
The acceptance window is an octagon corresponding to the projection of the  four-dimensional unit hypercube onto the perpendicular space. In this case, Eq.~(\ref{eq:ntf2}) gives:
\begin{widetext}
%
%
%%%%%%%%%%%%%%
\beq
|\widetilde{n}({\bf K}_\varparallel)|=\frac{N}{\lambda_+} \bigg|
\tfrac{\cos\big(\frac{\lambda_+ { K}_{\perp,1}  - { K}_{\perp,2}}{2} \big)}{K_{\perp,2}  ({ K}_{\perp,1} +K_{\perp,2} )}  -
\tfrac{\cos\big(\frac{\lambda_+ { K}_{\perp,1}  + { K}_{\perp,2}}{2} \big)}{K_{\perp,2}  ({ K}_{\perp,1} -K_{\perp,2} )}+
\tfrac{\cos\big(\frac{\lambda_+ { K}_{\perp,2}  - { K}_{\perp,1}}{2} \big)}{K_{\perp,1}  ({ K}_{\perp,2} +K_{\perp,1} )}-
\tfrac{\cos\big(\frac{\lambda_+ { K}_{\perp,2}  + { K}_{\perp,1}}{2} \big)}{K_{\perp,1}  ({ K}_{\perp,2} -K_{\perp,1} )}\bigg|, 
\label{eq:TF_octo}
\eeq
%%%%%%%%%%%%%%
%
%
\end{widetext}
for all reciprocal-lattice vector ${\bf K}$ with components $K_{\varparallel,j}={\bf K}. {\bf e}_{\varparallel,j}$ and $K_{\perp,j}={\bf K}. {\bf e}_{\perp,j}$. These expressions coincide with the one given in Ref.~\onlinecite{Janssen07}.

%
%
%%%%%%%%%%%%%%%%%%
\subsection{Structure factor}
%%%%%%%%%%%%%%%%%%
%
%

We are interested in computing the behavior of $S$ when $|{\bf K}_\varparallel|$ goes to $0$ for ${\bf K}_\varparallel \neq 0$. To this aim, we note that for a vector ${\bf K}=2\pi (s,t,u,v)$ [where $(s,t,u,v) \in \mathbb{Z}^4$], one has:
%
%
%%%%%%%%%%%%%%
\beqn
&&\lambda_+ K_{\perp,1} =   \frac{K_{\varparallel,1}}{\lambda_+} +2\pi(s-t+v),  \label{eq:trick_1}\\
&&\lambda_+ K_{\perp,2} =  \frac{K_{\varparallel,2}}{\lambda_+} +2\pi(-s-t+u),  \label{eq:trick_2}\\
&&K_{\perp,1}+ K_{\varparallel,1}  =  2\pi v,  \label{eq:trick_3}\\
&&K_{\perp,2}+ K_{\varparallel,2}  =  2\pi u.  \label{eq:trick_4}
\eeqn
%%%%%%%%%%%%%%
%
%
A close inspection of Eq.~(\ref{eq:TF_octo}) shows that one has to distinguish three different cases.\\

%
%
%%%%%%%%%%%%%%%%%%
\subsubsection{Symmetry axes: $S({\bf K}_\varparallel) {\sim} |{\bf K}_\varparallel|^4$}
%%%%%%%%%%%%%%%%%%
%
%

As can be seen in Eq.~(\ref{eq:TF_octo}), the denominator vanishes if one of the components $K_{\perp,i}=0$ or when $K_{\perp,1}=\pm K_{\perp,2}$. When $\bf{K}_{\perp}$ belongs to these four symmetry axes, the Fourier transform can be recast in a simple form. For simplicity, let us focus on the case where $K_{\perp,2}=0$ (the other cases being treated similarly) for which
%
%
%%%%%%%%%%%%%%
\beqn
|\widetilde{n}({\bf K}_\varparallel)|=\frac{N}{\lambda_+ K_{\perp,1}^2} 
\bigg| && 2\cos \left(\frac{K_{\perp,1}}{2}\right)- 2\cos \left(\frac{\lambda_+ K_{\perp,1}}{2}\right) + \nonumber \\
&& K_{\perp,1} \sin \left(\frac{\lambda_+ K_{\perp,1}}{2}\right) \bigg|.
\eeqn
%%%%%%%%%%%%%%
%
%
Using Eqs.~(\ref{eq:trick_1})-(\ref{eq:trick_3}), one then obtains  
%
%
%%%%%%%%%%%%%%
\beq
S({\bf K}_\varparallel)\underset{|{\bf K}_\varparallel| \to 0}{\simeq} \frac{N}{4 \lambda_+^4}  \bigg|\frac{K_{\varparallel,1}}{K_{\perp,1}} \bigg|^2.
\eeq
%%%%%%%%%%%%%%
%
%
As previously, to analyze the behavior of the structure factor for  small $|{\bf K}_\varparallel|$, we consider \mbox{${\bf K}^{(p)}=M^{p} {\bf K}$}. By construction, in the large-$p$ limit, the parallel components of ${\bf K}^{(p)}$ tend to zero as $\lambda_-^{p}$ and its perpendicular components diverge as $\lambda_+^p$. As a result, $S({\bf K}_\varparallel^{(p)})$ behaves as $\lambda_+^{-4p}$ so that, in this case, 
%
%
%%%%%%%%%%%%%%
\beq
S({\bf K}_\varparallel) \underset{|{\bf K}_\varparallel| \to 0}{\sim} |{\bf K}_\varparallel|^4.
\label{eq:scaling4}
\eeq
%%%%%%%%%%%%%%
%
%
This result actually holds for all $\bf{K}_{\perp}$ belonging to the four symmetry axes discussed above.

%
%
%%%%%%%%%%%%%%%%%%
\subsubsection{Generic cases: $S({\bf K}_\varparallel) {\sim} |{\bf K}_\varparallel|^8$ or $S({\bf K}_\varparallel) {\sim} |{\bf K}_\varparallel|^{12}$}
%%%%%%%%%%%%%%%%%%
%
%

When $\bf{K}_{\perp}$ does not belong to the four symmetry axes defined as $K_{\perp,1}=0$, $K_{\perp,2}=0$, and $K_{\perp,1}=\pm K_{\perp,2}$, one can again use Eqs.~(\ref{eq:trick_1}-\ref{eq:trick_4}) to express the structure factor as
%
%
%%%%%%%%%%%
\begin{widetext}
%
%
%%%%%%%%%%%%%%
\beqn
S({\bf K}_\varparallel)\underset{|{\bf K}_\varparallel| \to 0}{\simeq} \frac{N}{4 \lambda_+^4}  
\bigg|
\frac{(K_{\varparallel,1} K_{\perp,2}+K_{\varparallel,2} K_{\perp,1})(K_{\varparallel,1} K_{\perp,1}-K_{\varparallel,2} K_{\perp,2})}{K_{\perp,1} K_{\perp,2}({ K}_{\perp,1} +K_{\perp,2} )({ K}_{\perp,1}  -{ K}_{\perp,2})}
\bigg|^2,
\label{eq:leading}
\eeqn
%%%%%%%%%%%%%%
%
%
\end{widetext}
%%%%%%%%%%%
%
%
which leads to 
%
%
%%%%%%%%%%%%%%
\be
S({\bf K}_\varparallel) \underset{|{\bf K}_\varparallel| \to 0}{\sim} |{\bf K}_\varparallel|^8.
\label{eq:scaling8}
\ee
%%%%%%%%%%%%%%
%
%
However, as can be seen in Eq.~(\ref{eq:leading}), this leading contribution may vanish for some special ${\bf K}$. In this case, $S$ is given by the subleading contribution which gives 
%
%
%%%%%%%%%%%%%%
\be
S({\bf K}_\varparallel) \underset{|{\bf K}_\varparallel| \to 0}{\sim} |{\bf K}_\varparallel|^{12}.
\label{eq:scaling12}
\ee
%%%%%%%%%%%%%%
%
%

%
%
%%%%%%%%%%%%%%%%%%
\subsection{Integrated intensity function}
%%%%%%%%%%%%%%%%%%
%
%

To compute the integrated intensity function defined in Eq.~(\ref{eq:Zk}), we decompose the sum over all vector ${\bf K}_\varparallel$ with a norm smaller than $k$ as a sum over all quadruplets $(s,t,u,v)$ and their iterated vectors ${\bf K}_\varparallel^{(p)}=M^{p} {\bf K}_\varparallel$.  One can then write
%
%
%%%%%%%%%%%%%%
\beq
Z(k)=\frac{4\pi^2}{\mathcal A} \sum_{(s,t,u,v)} \sum_{p=p_{(s,t,u,v)}}^\infty S({\bf K}_\varparallel^{(p)}), 
\label{eq:Z_octo}
\eeq
%%%%%%%%%%%%%%
%
%
where $p_{(s,t,u,v)}$ is the smallest integer fulfilling the constraint \mbox{$|{\bf K}_\varparallel^{(p)}|<k$}. As discussed above, the behavior of $S$ in the large-$p$ (small-$|{\bf K}_\varparallel|$) limit, strongly depends on ${\bf K}_\varparallel$ [see Eqs.~(\ref{eq:scaling4}-\ref{eq:scaling12})]. This is in stark contrast with the Rauzy tiling and the Fibonacci chain where there is the same power-law scaling for all ${\bf K}_\varparallel$ [see Eqs.~(\ref{eq:scaling_S})-(\ref{eq:scaling_S_Fibo})]. 

However, since we are interested in the small-$k$ (large-$p$) limit, one only keeps the dominant terms in  Eq.~(\ref{eq:Z_octo}) that come from the symmetry axes and gives 
%
%
%%%%%%%%%%%%%%
\beqn
S({\bf K}_\varparallel^{(p)}) &\simeq&  |{\bf K}_\varparallel^{(p)}|^4 f(s,t,u,v), \label{eq:scaling_S_octo}\\
|{\bf K}_\varparallel^{(p)}| &\simeq&  \lambda_+^{-p} g(s,t,u,v). 
\label{eq:scaling_K_octo}
\eeqn
%%%%%%%%%%%%%%
%
%
We emphasize that, as for the Fibonacci chain, $f$ and $g$ are functions that do not depend on $p$, so that one straightforwardly gets  the following discrete scaling relation
%
%
%%%%%%%%%%%%
\be
Z(k/\lambda_+)=Z(k)/\lambda_+^{4}.
\ee 
%%%%%%%%%%%%
%
% 
The solution of this equation can be written as
%
%
%%%%%%%%%%%%
\be
Z(k)=k^4 F(\ln k/\ln \lambda_+),
\ee 
%%%%%%%%%%%%
%
% 
where $F(x+1)=F(x)$. As a result, $Z$ has a power-law envelope together with log-periodic oscillations.\\

%
%
%%%%%%%%%%%%%%%%%%%%%%%%%%%%%%%%%
%%%%%%%%%%%%%%%%%%%%%%%%%%%%%%%%%
\section{The Penrose tiling}
\label{app:Penrose}
%%%%%%%%%%%%%%%%%%%%%%%%%%%%%%%%%
%%%%%%%%%%%%%%%%%%%%%%%%%%%%%%%%%
%
%

The Penrose rhombus tiling \cite{Penrose79,DeBruijn81} can be built by CP from the five-dimensional hypercubic lattice $\mathbb{Z}^5$ (edge length $a=1$) along a well-known procedure (see, e.g., Ref.~\onlinecite{Duneau94} for details). 
For our purpose, let us consider the following orthogonal (non normalized) basis
%
%
%%%%%%%%%%%%%%
\beqn
 {\bf e}_{\varparallel,1}&=& \frac{2}{5} \left(1,{\rm c}_2,{\rm c}_4,{\rm c}_4,{\rm c}_2\right), \nonumber \\
 {\bf e}_{\varparallel,2}&=& \frac{2}{5} \left(0,{\rm s}_2,{\rm s}_4,-{\rm s}_4,-{\rm s}_2\right), \nonumber \\
 {\bf e}_{\perp,1}&=& \frac{2}{5} \left(1,{\rm c}_4,{\rm c}_2,{\rm c}_2,{\rm c}_4\right), \nonumber \\
 {\bf e}_{\perp,2}&=&  \frac{2}{5} \left(0,{\rm s}_4,-{\rm s}_2,{\rm s}_2,-{\rm s}_4\right),  \nonumber \\
{\bf e}_{\Delta}&=&\frac{1}{10}\left(1,1,1,1,1\right),
\eeqn
%%%%%%%%%%%%%%
%
%
that defines the three subspaces $E_\varparallel$, $E_\perp$ and $\Delta$. Here, we introduced the notation \mbox{${\rm c}_n=\cos(2\pi n /5)$} and \mbox{${\rm s}_n=\sin(2\pi n /5)$}. 
A point in $\mathbb{Z}^5$ is selected whenever it projects onto the perpendicular space $E_\perp + \Delta$ inside a three-dimensional acceptance window which is the projection of the five-dimensional unit hypercube onto this subspace. Remarkably,  selected points only fill five planes perpendicular to $\Delta$. Thus, the selection step only amounts to consider discrete sections of the acceptance window. Among all possible choices, the fivefold symmetric canonical Penrose tilings (known as star and sun \cite{Grunbaum87}) considered here correspond to the following sections:  one point which is the symmetry center of the tiling, two regular pentagons of side  $2 \sqrt{2/5}\cos(3\pi/10)$, two regular pentagons of side $  2  \, \tau \sqrt{2/5} \cos(3\pi/10)$, where $\tau=\frac{1+\sqrt{5}}{2}$ is the golden ratio.
%
%
%%%%%%%%%%%%%%%%%%
\subsection{Fourier transform}
%%%%%%%%%%%%%%%%%%
%
%

The Fourier transform of the tiling's vertices is obtained as a weighted sum of the Fourier transform of the four regular pentagons. For any vector $\mathbf{R}$ of the five-dimensional hypercubic lattice, the five-dimensional reciprocal lattices vectors $\mathbf{K}$ satisfy
\beq
\mathbf{K}\cdot \mathbf{R} = \mathbf{K}_\varparallel \cdot \mathbf{R}_\varparallel  + \mathbf{K}_\perp \cdot \mathbf{R}_\perp  + \mathbf{K}_\Delta \cdot \mathbf{R}_\Delta  = 2\pi\times \text{integer} .
\eeq

Then, after some algebra, Eq.~(\ref{eq:ntf2}) leads to:

\begin{widetext}
\beqn
|\tilde{n}(\mathbf{K}_\varparallel)| &=&N \bigg| \tfrac{8 ({\rm s}_2-{\rm s}_4)}{5 K_{\perp,2}} \bigg\{ \nonumber \\
&&
\tfrac{\cos({\rm c}_2 K_{\perp,1}+{\rm s}_2 K_{\perp,2}- 3 K_{\Delta})+\cos( K_{\perp,1}/2+({\rm s}_4+{\rm s}_2)K_{\perp,2}+ K_{\Delta})-\cos(K_{\perp,1}-3 K_{\Delta})-\cos(2{\rm c}_4 K_{\perp,1}- K_{\Delta})}{5K_{\perp,1}-(6{\rm s}_2+2{\rm s}_4) K_{\perp,2}}-\nonumber \\
&&
\tfrac{\cos({\rm c}_2 K_{\perp,1}-{\rm s}_2 K_{\perp,2}- 3 K_{\Delta})+\cos( K_{\perp,1}/2-({\rm s}_4+{\rm s}_2)K_{\perp,2}+ K_{\Delta})-\cos(K_{\perp,1}-3 K_{\Delta})-\cos(2{\rm c}_4 K_{\perp,1}- K_{\Delta})}{5K_{\perp,1}+(6{\rm s}_2+2{\rm s}_4) K_{\perp,2}}+\nonumber \\
&&
\tfrac{\cos({\rm c}_2 K_{\perp,1}-{\rm s}_2 K_{\perp,2}- 3 K_{\Delta})+\cos( K_{\perp,1}/2-({\rm s}_4+{\rm s}_2)K_{\perp,2}+ K_{\Delta})-\cos({\rm c}_4 K_{\perp,1}-{\rm s}_4 K_{\perp,2}-3 K_{\Delta})-\cos[(1+{\rm c}_2) K_{\perp,1}+{\rm s}_2 K_{\perp,2}- K_{\Delta}]}{5K_{\perp,1}-(6{\rm s}_4-2{\rm s}_2) K_{\perp,2}}-\nonumber \\
&&
\tfrac{\cos({\rm c}_2 K_{\perp,1}+{\rm s}_2 K_{\perp,2}- 3 K_{\Delta})+\cos( K_{\perp,1}/2+({\rm s}_4+{\rm s}_2)K_{\perp,2}+ K_{\Delta})-\cos({\rm c}_4 K_{\perp,1}+{\rm s}_4 K_{\perp,2}-3 K_{\Delta})-\cos[(1+{\rm c}_2) K_{\perp,1}-{\rm s}_2 K_{\perp,2}- K_{\Delta}]}{5K_{\perp,1}+(6{\rm s}_4-2{\rm s}_2) K_{\perp,2}}\bigg\}\bigg|,\nonumber \\
\label{eq:Sq_penrose}
\eeqn
\end{widetext}
for all reciprocal-lattice vector ${\bf K}$ with components $K_{\varparallel,j}={\bf K}. {\bf e}_{\varparallel,j}$, $K_{\perp,j}={\bf K}. {\bf e}_{\perp,j}$, and $K_{\Delta}={\bf K}. {\bf e}_{\Delta}$.

%
%
%%%%%%%%%%%%%%%%%%
\subsection{Structure factor}
%%%%%%%%%%%%%%%%%%
%
%

We are interested in computing the behavior of $S$ when $|{\bf K}_\varparallel|$ goes to $0$ for ${\bf K}_\varparallel \neq 0$. To this aim, we note that for a vector ${\bf K}=2\pi (s,t,u,v,w)$ [where $(s,t,u,v,w) \in \mathbb{Z}^5$], one has
\begin{widetext}
%
%
%%%%%%%%%%%%%%
\beqn
&&{\rm c}_2 K_{\perp,1}+{\rm s}_2 K_{\perp,2}- 3 K_{\Delta} = -{\rm c}_4 K_{\varparallel,1}+{\rm s}_4 K_{\varparallel,2}- 5 K_{\Delta} +2\pi v,  \label{eq:trickp_1}\\
&&{\rm c}_2 K_{\perp,1}- {\rm s}_2 K_{\perp,2}- 3 K_{\Delta} =   -{\rm c}_4 K_{\varparallel,1}-{\rm s}_4 K_{\varparallel,2}- 5 K_{\Delta} +2\pi u, \label{eq:trickp_2}\\
&&K_{\perp,1}/2+({\rm s}_4+{\rm s}_2)K_{\perp,2}+ K_{\Delta} = -K_{\varparallel,1}/2+({\rm s}_4-{\rm s}_2)K_{\varparallel,2}+ 5 K_{\Delta}  -2\pi (u+w), \label{eq:trickp_3}\\
&&K_{\perp,1}/2-({\rm s}_4+{\rm s}_2)K_{\perp,2}+ K_{\Delta} =  -K_{\varparallel,1}/2-({\rm s}_4-{\rm s}_2)K_{\varparallel,2}+ 5 K_{\Delta}  -2\pi (v+t),  \label{eq:trickp_4}\\
&&K_{\perp,1}-3 K_{\Delta} = -K_{\varparallel,1}-5 K_{\Delta} +2\pi s, \label{eq:trickp_5}\\
&&2{\rm c}_4 K_{\perp,1}- K_{\Delta} =  -2{\rm c}_2 K_{\varparallel,1}-5 K_{\Delta} +2\pi (t+w), \label{eq:trickp_6}\\
&& {\rm c}_4 K_{\perp,1}+{\rm s}_4 K_{\perp,2}-3 K_{\Delta} = -{\rm c}_2 K_{\varparallel,1}-{\rm s}_2 K_{\varparallel,2}-5 K_{\Delta} +2\pi t, \label{eq:trickp_7}\\
&& {\rm c}_4 K_{\perp,1}-{\rm s}_4 K_{\perp,2}-3 K_{\Delta} = -{\rm c}_2 K_{\varparallel,1}+{\rm s}_2 K_{\varparallel,2}-5 K_{\Delta}+2\pi w, \label{eq:trickp_8}\\
&&(1+{\rm c}_2) K_{\perp,1}+{\rm s}_2 K_{\perp,2}- K_{\Delta} = -(1+{\rm c}_4) K_{\varparallel,1}+{\rm s}_4 K_{\perp,2}- 5 K_{\Delta} +2\pi (s+v),\label{eq:trickp_9}\\
&&(1+{\rm c}_2) K_{\perp,1}-{\rm s}_2 K_{\perp,2}- K_{\Delta} = -(1+{\rm c}_4) K_{\varparallel,1}-{\rm s}_4 K_{\perp,2}- 5 K_{\Delta} +2\pi (s+u).\label{eq:trickp_10}
\eeqn
%%%%%%%%%%%%%%
%
%
\end{widetext}

Keeping in mind that  $5 K_{\Delta}=\pi (s+t+u+v+w)$, one can finally rewrite Eq.~(\ref{eq:Sq_penrose}) as a function of $K_{\varparallel,j}$ and  $K_{\perp,j}$ only. 
In the limit  where $|{\bf K}_\varparallel| \to 0$ one then gets generically
\begin{widetext}
%
%
%%%%%%%%%%%%%%
\beq
S({\bf K}_\varparallel) \underset{|{\bf K}_\varparallel| \to 0}{\simeq} N (\sqrt{5}-2)^2 \Bigg|\frac{(K_{\perp,1}^2+ K_{\perp,2}^2)[(K_{\varparallel,1}^2-K_{\varparallel,2}^2)K_{\perp,2}-2K_{\varparallel,1} K_{\varparallel,2} K_{\perp,1}]}{K_{\perp,2}(5K_{\perp,1}^4-10 K_{\perp,1}^2 K_{\perp,2}^2+K_{\perp,2}^4)} \Bigg|^2.
\eeq
%%%%%%%%%%%%%%
%
%
\end{widetext}

However, when one of the denominators in Eq.~(\ref{eq:Sq_penrose}) vanishes, one gets different expressions that are easily obtained along the same line. %

To analyze the behavior of the structure factor for small $|{\bf K}_\varparallel|$, we consider the matrix
\beq
M=\left(\begin{array}{ccccc}0&1&0&0&1\\
1&0&1&0&0\\
0&1&0&1&0\\
0&0&1&0&1\\
1&0&0&1&0
\end{array}
\right),
\eeq
whose  eigenspaces are $E_\varparallel$, $E_\perp$ and $E_\Delta$ with eigenvalues $\lambda_\varparallel=1/\tau$, $\lambda_\perp=-\tau$ and $\lambda_\Delta=2$, respectively. By construction, in the large-$p$ limit, the parallel components of ${\bf K}^{(p)}=M^{p} {\bf K}$ vanishes as $\lambda_\varparallel^{p}$ whereas its perpendicular components diverge as $\lambda_\perp^{p}$. As a result, in the large-$p$ limit, $S({\bf K}_\varparallel^{(p)})$ behaves as $\lambda_\varparallel^{8p}$ so that
%
%
%%%%%%%%%%%%%%
\be
S({\bf K}_\varparallel) \underset{|{\bf K}_\varparallel| \to 0}{\sim} |{\bf K}_\varparallel|^8.
\label{eq:scaling8}
\ee
%%%%%%%%%%%%%%
%
%

%
%
%%%%%%%%%%%%%%%%%%
\subsection{Integrated intensity function}
%%%%%%%%%%%%%%%%%%
%
%

As  for the octagonal tiling, we decompose the sum over ${\bf K}_\varparallel$ in the integrated intensity function defined in Eq.~(\ref{eq:Zk}) as a sum over all quintuplets $(s,t,u,v,w)$ and their iterated under ${\bf K}_\varparallel^{(p)}=M^{p} {\bf K}_\varparallel$.  One can then write
%
%
%%%%%%%%%%%%%%
\beq
Z(k)=\frac{4\pi^2}{\mathcal A} \sum_{(s,t,u,v,w)} \sum_{p=p_{(s,t,u,v,w)}}^\infty S({\bf K}_\varparallel^{(p)}), 
\label{eq:Z_penrose}
\eeq
%%%%%%%%%%%%%%
%
%
where $p_{(s,t,u,v,w)}$ is the smallest integer fulfilling the constraint \mbox{$|{\bf K}_\varparallel^{(p)}|<k$}. 
In the small-$k$ (large-$p$) limit , one can check that
%
%
%%%%%%%%%%%%%%
\beq
S({\bf K}_\varparallel^{(p)}) \simeq  |{\bf K}_\varparallel^{(p)}|^8 f(s,t,u,v,w),
\label{eq:scaling_K_penrose}
\eeq
%%%%%%%%%%%%%%
%
%
for any quintuplet $(s,t,u,v,w)\in \mathbb{Z}^5$. However, contrary to the octagonal tiling, the scaling of ${\bf K}_\varparallel^{(p)}$ with $p$ depends on the quintuplet. For quintuplets that do not annihilate the denominator in Eq.~(\ref{eq:Sq_penrose}), one gets:
%
%
%%%%%%%%%%%%%%
\beq
|{\bf K}_\varparallel^{(p)}| \simeq  \tau^{-p} g(s,t,u,v,w), 
\eeq \\
%%%%%%%%%%%%%%
%
%
or, in other words, \mbox{$|{\bf K}_\varparallel^{(p)}/{\bf K}_\varparallel^{(p+1)}|=\tau$}.
Importantly, when one of the denominators in Eq.~(\ref{eq:Sq_penrose}) vanishes, one gets a weaker relation since one only has \mbox{$|{\bf K}_\varparallel^{(p)}/{\bf K}_\varparallel^{(p+2)}|=\tau^2$}. As a direct consequence, one gets the following discrete scaling relation
%
%
%%%%%%%%%%%%
\be
Z(k/\tau^2)=Z(k)/\tau^{16}.
\ee 
%%%%%%%%%%%%
%
% 
The solution of this equation can be written as
%
%
%%%%%%%%%%%%
\be
Z(k)=k^8 F(\ln k/\ln \tau^2),
\ee 
%%%%%%%%%%%%
%
% 
where $F(x+1)=F(x)$. As a result, $Z$ has a power-law envelope together with log-periodic oscillations.

%\bibliography{./biblio_hyperuniformity}


\end{document}